\documentclass[aps,twocolumn,pra,superscriptaddress,showpacs,tightenlines]{revtex4}
\usepackage{mathrsfs}
\usepackage{amssymb}
\usepackage{amsmath}
\usepackage{graphicx}
\usepackage{epsfig}
\usepackage{txfonts}
\usepackage{subfigure}
\usepackage{amsfonts}

\input{tcilatex}

\begin{document}

\title{Information gain versus interference in Bohr's principle of
complementarity}
\author{Yan Liu}
\affiliation{Key Laboratory of Low-Dimensional Quantum Structures and Quantum Control of
Ministry of Education, Department of Physics and Synergetic Innovation
Center for Quantum Effects and Applications, Hunan Normal University,
Changsha 410081, China}
\affiliation{Center for Emergent Matter Science, RIKEN, Saitama 351-0198, Japan}
\author{Jing Lu}
\email{lujing@hunnu.edu.cn}
\affiliation{Key Laboratory of Low-Dimensional Quantum Structures and Quantum Control of
Ministry of Education, Department of Physics and Synergetic Innovation
Center for Quantum Effects and Applications, Hunan Normal University,
Changsha 410081, China}
\affiliation{Center for Emergent Matter Science, RIKEN, Saitama 351-0198, Japan}
\author{Lan Zhou}
\affiliation{Key Laboratory of Low-Dimensional Quantum Structures and Quantum Control of
Ministry of Education, Department of Physics and Synergetic Innovation
Center for Quantum Effects and Applications, Hunan Normal University,
Changsha 410081, China}
\affiliation{Center for Emergent Matter Science, RIKEN, Saitama 351-0198, Japan}
\author{Franco Nori}
\affiliation{Center for Emergent Matter Science, RIKEN, Saitama 351-0198, Japan}
\affiliation{Physics Department, The University of Michigan, Ann Arbor, Michigan 48109-1040, USA}
\date{\today }

\begin{abstract}
Based on modern quantum measurement theory, we use Zurek's ``triple
model" to study, from the viewpoint of quantum information theory,
the wave and particle nature of a photon in a symmetric Mach-Zehnder
interferometer. In the process of quantum measurement, the state of
both the system and the detector is not an entangled state but a
correlated state. We find that the information gain about the photon
is related to the correlations (including classical and quantum
correlations) between the photon and the detector. We also derive
the relationship between the information gain and the fringe
visibility. We find that the classical correlations remain
consistent with the path distinguishability and can be used to
describe the particle-like property of the photon. Quantum
correlations are not exactly the same as fringe visibility, but both
can represent the quantum coherence of the photon. Finally, we
provide an analytical expression for quantum correlations of one
type of two-qubit separable states.

\end{abstract}

\maketitle


\narrowtext


\section{Introduction}

Unlike classical particles, a quantum system behaves either as a
particle or as a wave. This property is called wave-particle
duality, which is one of the famous intriguing feature of quantum
mechanics. Quantum properties which are equally real but mutually
exclusive are called complementary \cite{Bohr1928a,Bohr1928b}.
Wave-particle duality is well described in Bohr's complementarity
principle, which is sometimes phrased as follows: waves and
particles are two distinct types of complementarity in nature, and
the experimental situation determines the particle or wave nature of
a quantum system; however, the simultaneous observation of wave and
particle behavior is impossible. Mutual exclusiveness is regarded by
Bohr as a ``necessary'' element in the complementarity principle to
ensure its inner consistency~\cite{FP22(92)1435}. The usual
discussion about wave-particle duality starts from a physical system
with two alternatives, typically, a two-way interferometer such as
Young's double-slit experiment or a Mach-Zehnder setup. If one
performs quantum measurements to determine which way a quantum
particle is taken (particle-like property), the interference pattern
(wave-like property) is partially or completely destroyed by the
partial or complete knowledge of the ``which-way" information. The
more one obtains the which-way information, the more the loss of
interference~\cite{PRD19(79)473,PLA128(88)391,Jaeger1995,PRL77(96)2154,Bliokh2013}.
Many experiments have demonstrated this complementarity with
different quantum systems, like atoms~\cite{Durr1998},
lasers~\cite{Schwindt1999}, nuclear magnetic
resonance~\cite{Peng2005}, and single
photons~\cite{Jacques2007a,Jacques2007b,Kocsis2011,Yan2015}.
Obviously, the concept of measurement plays an important role in a
logically consistent description of the wave-particle properties in
a two-way interferometer.

Bohr's interpretation of the interferometry experiments invokes the
concept of wave-function collapse. Measurements perturb the wave
function, so the collapse hypothesis is responsible for mutually
exclusive quantities. Based on classical concepts and intuitions,
Bohr thought that the stages of preparation and registering the
quantum objects require classical apparatuses, which draw an obvious
border between the quantum and the classical world. However, this
concept is contrary to the belief that quantum theory is universally
applicable and classical reality may be reconstructed or
reconstituted from quantum dynamics. By treating the apparatus as a
large number of particles or a large number of degrees of freedom
obeying the Sch\"{o}dinger equation, the loss of interference is
explained by the nonseparable correlation between the measuring
apparatus and the system being observed,
where the information of the measured system is stored in the pointer states~%
\cite{PRD24(81)1516,Brasil2015} of the apparatus. Such a correlation between two systems
is called entanglement. Mathematically, the loss of interference is described by the elimination
of the off-diagonal elements of the system's reduced density matrix, which
is called decoherence in the terminology of quantum measurement theory. By
utilizing entanglement, the position-momentum uncertainty relation is found
to play no role in the principle of complementarity~\cite{PRL77(96)2154,351(91)111}. Note that a quantitative formulation of Bohr's
complementarity principle can be derived from the position-momentum uncertainty~%
\cite{PRD19(79)473,Storey1995}. To find out what makes the quantum apparatus to have a
number of states equal to a number of possible distinct outcomes of the
measurement, an environment is necessarily introduced to interact with the quantum
apparatus~\cite{PRD26(82)1862} for the interferometry experiments.

In the duality relation~\cite{PRL77(96)2154}, the wave nature is
described in terms of the visibility of the interference pattern.
The particle nature is characterized by the path distinguishability,
which is a measure of the which-way information. The wave-particle
duality is demonstrated by the visibility of fringes setting limits
on the which-way information. With the advent of quantum information
theory, there are many measurements which can quantify how much
information about a system A is stored inside the other system B,
including both von Neumann~\cite{Milburn2010} and weak
measurements~\cite{Aharonov1988,Kofman2012,Dressel2014}. Classical
correlations (CC)~\cite{Vedral2001,Vedral2003} and Quantum discord
(QD)~\cite{PRL88(01)017901,Vedral2012} are based on von Neumann
measurements and are originally introduced as an
information-theoretic approach to decoherence mechanisms in a
quantum measurement process. The close link between the measurement
process and quantitative analysis of the complementarity suggests
that CC (QD) and which-way information (visibility) might be
different views of one single phenomenon. In this paper, we
investigate the relationship between the amount of information that
can be extracted about the measurement process and interference in a
which-way experiment.

This paper is organized as follows. In Sec.~\ref{Sec:2}, we briefly
review the well-known method of quantifying wave--particle duality.
In Sec.~\ref {Sec:3}, we study the amount of information gain from
the quantum measurement process by using modern quantum measurement
theory in a which-way experiment. The relationship between the
amount of information gain and the fringe visibility are also
investigated. Finally, We conclude this work in Sec.~\ref{Sec:4}.


\section{\label{Sec:2}wave--particle duality relation}

\begin{figure}[tbp]
\includegraphics [bb=-60 429 459 679, width=3 in]{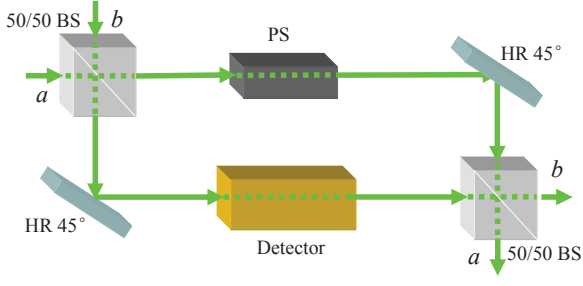}
\caption{(Color online). Schematic of a symmetric Mach-Zehnder
interferometer. Here, $PS$ refers to phase shifter, $PS$ to beam
splitters, and $HR$ to high reflector.} \label{fig1.eps}
\end{figure}

A symmetric Mach-Zehnder interferometer has two 50:50 beam splitters
(BSs) and a phase shifter (PS) as shown in Fig.~\ref{fig1.eps}.
Between the two BSs, two possible routes $a$ and $b$ are
macroscopically well separated, which are represented by orthogonal
unit vectors $|a\rangle $, $|b\rangle$ of a two dimensional Hilbert
space. Hereafter, states $|a\rangle $ and $|b\rangle $ are called
path states.
The wave-function of a photon incident on path $a$ ($b$) is changed to an equally weighted superposition
\begin{subequations}
\label{II-01}
\begin{eqnarray}
\left\vert a\right\rangle &\rightarrow &\frac{1}{\sqrt{2}}\left( \left\vert
a\right\rangle +\left\vert b\right\rangle \right) \\
\left\vert b\right\rangle &\rightarrow &\frac{1}{\sqrt{2}}\left( \left\vert
a\right\rangle -\left\vert b\right\rangle \right)
\end{eqnarray}%
by the first BS. As a single photon propagates along this path, a
relative phase $\phi $ is accumulated between states $|a\rangle $
and $|b\rangle $. To obtain the knowledge of the actual path a
photon has taken, a which-way detector (WWD) is introduced, which
performs a quantum non-demolition measurement without any backaction
on this photon. If a photon propagates along a path $a$, the initial
state $\rho _{\rm in}^{D}$ of the WWD remains unchanged; however,
$\rho _{\rm in}^{D}$ is changed to $U\rho _{\rm in}^{D}U^{\dag }$ if
the photon propagates on path $b$. States $\rho _{\rm in}^{D}$ and
$U\rho_{\rm in}^{D}U^{\dag }$ are not always orthogonal. By defining
the operators $\sigma _{\alpha}$ ($\alpha =x,y,x$) in terms of the
states $|a\rangle $ and $|b\rangle $, e.g., $\sigma _{z}=\left\vert
a\right\rangle \left\langle a\right\vert -\left\vert b\right\rangle
\left\langle b\right\vert $, the degree of freedom described by the
path states is analogous to a spin. Therefore, the initial state of
a photon is generally characterized by the density matrix
\end{subequations}
\begin{equation}
\rho _{\rm in}^{Q}=\frac{1}{2}\left( 1+S_{x}\sigma _{x}+S_{y}\sigma
_{y}+S_{z}\sigma _{z}\right)  \label{II-02}
\end{equation}
with an initial Bloch vector $\vec{S}=\left(
S_{x},S_{y},S_{z}\right) $. The initial product state $\rho _{\rm
in}^{Q}\otimes \rho _{\rm in}^{D}$ of the two subsystems is evolved
into
\begin{eqnarray}
\rho _{f} &=&\frac{1}{4}\left( 1-S_{x}\right) \left( 1+\sigma _{x}\right)
\otimes \rho _{\rm in}^{D}  \notag \\
&&+\frac{1}{4}\left( 1+S_{x}\right) \left( 1-\sigma _{x}\right) \otimes
U\rho _{\rm in}^{D}U^{\dagger }  \notag \\
&&-\frac{1}{4}e^{-i\phi }\left( S_{z}-iS_{y}\right) \left( \sigma
_{z}-i\sigma _{y}\right) \otimes \rho _{\rm in}^{D}U^{\dagger }  \notag \\
&&-\frac{1}{4}e^{i\phi }\left( S_{z}+iS_{y}\right) \left( \sigma
_{z}+i\sigma _{y}\right) \otimes U\rho _{\rm in}^{D},  \label{II-03}
\end{eqnarray}%
after the photon has gone through the Mach-Zehnder interferometer,
which establishes the correlation between the photon and the WWD.
Then, the which-way information is stored in the WWD. The state of
the single photon reads
\begin{eqnarray}
\rho _{f}^{Q} &=&\frac{1}{4}\left( 1-S_{x}\right) \left( 1+\sigma
_{x}\right) +\frac{1}{4}\left( 1+S_{x}\right) \left( 1-\sigma _{x}\right)
\notag \\
&&-\frac{1}{4}e^{-i\phi }\left( S_{z}-iS_{y}\right) \left( \sigma
_{z}-i\sigma _{y}\right) {\rm Tr}_{D}\left( \rho _{\rm in}^{D}U^{\dagger }\right)
\notag \\
&&-\frac{1}{4}e^{i\phi }\left( S_{z}+iS_{y}\right)\left( \sigma
_{z}+i\sigma _{y}\right) {\rm Tr}_{D}\left( U\rho _{\rm in}^{D}\right)  \label{II-04}
\end{eqnarray}%
by tracing over the degrees of freedom of the WWD. The probability for the photon emerging from the output $a$
\begin{eqnarray}
P^{a} &=&{\rm Tr}_{Q}\left[ \frac{1}{2}\left( 1+\sigma _{z}\right) \rho _{f}^{Q}%
\right]  \notag \\
&=&\frac{1}{2}-\frac{1}{2}\sqrt{S_{y}^{2}+S_{z}^{2}}\left\vert {\rm Tr}_{D}\left(
U\rho _{\rm in}^{D}\right) \right\vert \cos \left( \alpha +\beta +\phi \right)
\label{II-05}
\end{eqnarray}%
is used to define the visibility
\begin{equation}
V\equiv \frac{P_{\max }^{a}-P_{\min }^{a}}{P_{\max }^{a}+P_{\min }^{a}}=%
\sqrt{S_{y}^{2}+S_{z}^{2}}\left\vert {\rm Tr}_{D}\left( U\rho
_{\rm in}^{D}\right) \right\vert  \label{II-06}
\end{equation}%
of the interference pattern, where the constant phase shifts $\alpha $ and $\beta $ are the phases of $S_{z}+iS_{y}$ and ${\rm Tr}_{D}\left( U\rho _{in}^{D}\right) $
respectively. To extract the information in the final state of the WWD%
\begin{equation}
\rho _{f}^{D}=\frac{1-S_{x}}{2}\rho _{\rm in}^{D}+\frac{1+S_{x}}{2}U\rho
_{\rm in}^{D}U^{\dag },  \label{II-07}
\end{equation}%
an observable must be chosen for the readout. Englert~\cite{PRL77(96)2154} introduced the
distinguishability
\begin{equation}
D={\rm Tr}_{D}\left\vert \frac{1-S_{x}}{2}\rho _{\rm in}^{D}-\frac{1+S_{x}}{2}%
U\rho _{\rm in}^{D}U^{\dag }\right\vert   \label{II-08}
\end{equation}
to be the maximum of the difference of probabilities of the correct
and incorrect decisions about the paths. Then the fringe visibility
and the maximum amount of which-way information are bound in a
trade-off relation
\begin{equation}
D^{2}+\frac{1-P^{2}}{V_{0}^{2}}V^{2}\leq 1,  \label{II-09}
\end{equation}
where $P=\left\vert S_{x}\right\vert $ is the predictability and
$V_{0}=\sqrt{S_{y}^{2}+S_{z}^{2}}$ is a priori fringe visibility.
Actually, the parameters $P$ and $V_{0}$ construct a trade-off
relation $P^{2}+V_{0}^{2}\leq 1$, which is known as the duality
relationship for preparation~\cite{PLA128(88)391}.

Special attention is paid in Ref.~\cite{PRL77(96)2154} on the
initial state with $S_{x}=0$ and $S_{z}+iS_{y}=e^{-i\theta }$ (in
this case, $P=0$, $V_{0}=1$) to emphasize the quantum properties of
the WWD, which enforce duality and make sure that the principle of
complementarity is not circumvented. In this sense, we will set
$S_{x}=0$ and $S_{z}+iS_{y}=e^{-i\theta }$ in the rest of our paper.

\section{\label{Sec:3}information gain versus interference}
\subsection{Traditional approach}
In order to acquire the which-way information, a WWD is introduced
to interact with the photon, and a suitable observable must be
chosen for the readout. For simplicity, we take the initial state of
the WWD as a pure state, i.e., $\rho_{\rm in}^{D}=\left\vert
d\right\rangle \left\langle d\right\vert $.
Reference~\cite{PRL77(96)2154} relies on von Neumann's notion of
quantum measurement, i.e., the WWD interacts with the photon and
becomes entangled with it, and the state of the combined
photon-detector system can then be written as
\begin{eqnarray}
\label{III-01}
\left\vert\Psi\right\rangle&=&\frac{1}{\sqrt{2}}\left( \left\vert
a\right\rangle  \otimes \left\vert d\right\rangle + \left\vert
b\right\rangle  \otimes U\left\vert d\right\rangle \right).
\end{eqnarray}
Next, the ``likelihood for guessing the way right'' is
introduced~\cite{PRL77(96)2154} to describe the which-way
information. And the largest amount of information can be obtained
when the eigenstates of the observable are also the eigenstates of
the difference $\left\vert d\right\rangle \left\langle
d\right\vert-U\left\vert d\right\rangle \left\langle d\right\vert
U^{\dagger }$. From Eqs.~(\ref {II-06}, \ref {II-08}), the
visibility of the interference patten $V=\left|\left\langle
d\right\vert U\left\vert d\right\rangle\right|$, and the path
distinguishability $D=\sqrt{1-\left|\left\langle d\right\vert
U\left\vert d\right\rangle\right|^{2}}$, which satisfy the
complementarity principle $V^{2}+D^{2}=1$. We note that the above
strategy is equal to the theory of quantum state discrimination with
minimum error \cite{Helstrom,Chefles2000}. The error-minimum
discrimination is described by the projection operators $\Pi
_{A}=\left\vert M_{A}\right\rangle \left\langle M_{A}\right\vert $
and $ \Pi_{B}=\left\vert M_{B}\right\rangle \left\langle
M_{B}\right\vert $, where optimum measurement vectors are given by
\begin{eqnarray}
\label{III-02}
\left\vert M_{A}\right\rangle &=&\frac{1}{m}\sqrt{\frac{1+m}{2}}
\left\vert d\right\rangle -e^{i\varphi}\frac{1}{m}\sqrt{\frac{1-m}{2}}U\left\vert d\right\rangle ,  \notag \\
\left\vert M_{B}\right\rangle &=& e^{-i\varphi}\frac{1}{m}\sqrt{\frac{1-m}{2}}
\left\vert d\right\rangle -\frac{1}{m}\sqrt{\frac{1+m}{2}} U\left\vert d\right\rangle
\end{eqnarray}
with $m=\sqrt{1-V^{2}}$, and $\varphi$ being the relative phase. In
fact, the optimum measurement vectors $\left\vert
M_{A}\right\rangle$ and $\left\vert M_{B}\right\rangle$ are exactly
the eigenstates of the difference $\left\vert d\right\rangle
\left\langle d\right\vert-U\left\vert d\right\rangle \left\langle
d\right\vert U^{\dagger}$.
\subsection{Classical correlations versus interference}
Here we try to evaluate the which-way information on a different
point of view with the aid of quantum information theory. In quantum
information theory, the classical correlations between the photon
and the WWD can be captured by
\begin{eqnarray}
\label{III-03}
\mathcal{J}(\rho'_{f})=\max \left[S(\rho
^{Q})- S(\rho ^{Q}|\{{\Pi _{k}\}})\right],
\end{eqnarray}
where $\rho^{Q}$ is the reduced density operator for the photon,
$S(\rho^{Q})$ is the von Neumann entropy, $S(\rho ^{Q}|\{{\Pi
_{k}\}})$ is the quantum conditional entropy, and $\{\Pi _{k}\}$ is
a set of projectors performed locally on the WWD. To calculate the
quantum conditional entropy, we choose the von Neumann projection,
and the orthogonal projection operators $\Pi _{1}=\left\vert
M_{1}\right\rangle \left\langle M_{1}\right\vert $ and $
\Pi_{2}=\left\vert M_{2}\right\rangle \left\langle M_{2}\right\vert
$, where $\left\vert M_{1}\right\rangle $ and $\left\vert
M_{2}\right\rangle$ are given by
\begin{eqnarray}
\label{III-04}
\left\vert M_{1}\right\rangle &=&\frac{\sin \gamma }{\sqrt{1-V^{2}}}%
\left\vert d\right\rangle +e^{i\varphi }\left( \cos \gamma -\sin \gamma \frac{V}{\sqrt{%
1-V^{2}}}\right) U\left\vert d\right\rangle ,  \notag \\
\left\vert M_{2}\right\rangle &=&e^{-i\varphi }\frac{\cos \gamma }{\sqrt{1-V^{2}}}%
\left\vert d\right\rangle -\left( \sin \gamma +\cos \gamma \frac{V}{\sqrt{%
1-V^{2}}}\right) U\left\vert d\right\rangle   \notag \\
&&
\end{eqnarray}
with $V=\left|\left\langle d\right\vert U\left\vert
d\right\rangle\right| $, $0\leq \varphi \leq 2\pi $ and $0\leq
\gamma \leq 2\pi$. Actually, our strategy to discriminate two
non-orthogonal states of the WWD and obtain the which-way
information is considering a set of von Neumann projection performed
locally on the WWD. The CC is the largest classical mutual
information gained about photon after a measurement of the WWD.

(1) \emph{Based on von Neumann's notion of quantum measurement}

\begin{figure}[tbp]
\includegraphics [clip=true,height=4.5cm,width=8cm]{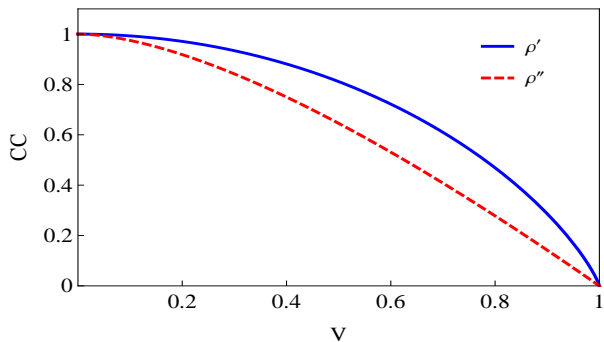}
\caption{(Color online). The relationship between the classical
correlations and the fringe visibility $V$. } \label{fig2.eps}
\end{figure}

The state of the combined photon-detector system in Eq.~(\ref {III-01}) can be written as
\begin{equation}
\label{III-05} \rho^{
\prime}=\left\vert\Psi\right\rangle\left\langle\Psi\right\vert
=\frac{1}{2}\left(
\begin{array}{cccc}
1 & 0 & V & \sqrt{1-V^{2}} \\
0 & 0 & 0 & 0 \\
V & 0 & V^{2} & V\sqrt{1-V^{2}} \\
\sqrt{1-V^{2}} & 0 & V\sqrt{1-V^{2}} & 1-V^{2}
\end{array}
\right)
\end{equation}
in the basis $\left\{\left\vert a\right\rangle \left\vert
d\right\rangle,\left\vert a\right\rangle \left\vert
d_{\perp}\right\rangle,\left\vert b\right\rangle \left\vert
d\right\rangle,\left\vert b\right\rangle \left\vert
d_{\perp}\right\rangle\right\}$. After a straightforward
calculation, we can obtain the CC
\begin{eqnarray}
\label{III-06} \mathcal{J}(\rho^{ \prime})&=&-\frac{1+V}{2}\log
\left(\frac{1+V}{2}\right)-\frac{1-V}{2}\log \left(
\frac{1-V}{2}\right),
\end{eqnarray}
which is shown by the blue-solid curve in Fig.~\ref{fig2.eps}.

(2) \emph{Based on Zurek's ``triple model" of quantum measurement}

To realize the wave-function collapse of the measured system by
establishing an entanglement between the system and the apparatus in
quantum measurement theory, an observer must first select the state
of detector and then read it out. To avoid this subjective
selection, Zurek introduced a ``triple model" of quantum measurement
process, which consists of a measured system (photon), an apparatus
(WWD), and an environment. According the environment-induced
superselection, the entangled state of the combined photon-detector
system becomes a correlated state
\begin{eqnarray}
\label{III-07} \rho^{\prime \prime}&=&\frac{1}{2}\left( \left\vert
a\right\rangle \left\langle a\right\vert \otimes \left\vert
d\right\rangle \left\langle d\right\vert+ \left\vert b\right\rangle
\left\langle b\right\vert \otimes U\left\vert d\right\rangle
\left\langle d\right\vert U^{\dagger }\right),
\end{eqnarray}
which can be written as
\begin{equation}
\label{III-08} \rho^{\prime \prime} =\frac{1}{2}\left(
\begin{array}{cccc}
1 & 0 & 0 & 0 \\
0 & 0 & 0 & 0 \\
0 & 0 & V^{2} & V\sqrt{1-V^{2}} \\
0 & 0 & V\sqrt{1-V^{2}} & 1-V^{2}
\end{array}
\right)
\end{equation}
in the basis $\left\{\left\vert a\right\rangle \left\vert
d\right\rangle,\left\vert a\right\rangle \left\vert
d_{\perp}\right\rangle,\left\vert b\right\rangle \left\vert
d\right\rangle,\left\vert b\right\rangle \left\vert
d_{\perp}\right\rangle\right\}$. It can be found that the state
$\rho^{\prime \prime}$ is no longer an entangled state, and the
state $\left\vert d\right\rangle $ ($U\left\vert d\right\rangle $)
becomes correlated with $\left\vert a\right\rangle $ ($\left\vert
b\right\rangle $). We note that the states $\left\vert
d\right\rangle$ and $U\left\vert d\right\rangle $ are two pointer
states of the WWD. If the state of the WWD is $\left\vert
d\right\rangle$ ($U\left\vert d\right\rangle $), the observer can
infer that the photon passes through the path $a$ ($b$). Note that
the observer just reads out the pointer states of WWD. If
$\left\vert d\right\rangle =U\left\vert d\right\rangle $, the two
pointer states of the WWD are the same. Two equal pointer states
cannot indicate two possible outcomes. Therefore, the paths of the
photon cannot be distinguished, and we cannot obtain the which-way
information, i.e., $D=0$. If $\left\vert d\right\rangle $ and
$U\left\vert d\right\rangle $ are mutually orthogonal, the two
pointer states are completely different and can be perfectly
discriminated. Two different pointer states correspond to two
different paths of the photon, and a perfect distinction of the
photon path is achieved when $D=1$. However, when $\left\vert
d\right\rangle $ and $U\left\vert d\right\rangle$ are not mutually
orthogonal, it is impossible to discriminate them perfectly.



 According to the Eq.~(\ref {III-08}), after a straightforward calculation, we obtain
\begin{widetext}
\begin{eqnarray}
\label{III-09}
S(\rho ^{Q})- S(\rho ^{Q}|\{{\Pi
_{k}\}})&=&1+\frac{1}{2}\left[ \cos ^{2}\gamma \right] \log \left[
\cos ^{2}\gamma \right] +
\frac{1}{2}\left[\sin ^{2}\gamma \right] \log \left[ \sin ^{2}\gamma \right] \notag\\
&&+\frac{1}{2}\left[  \left( \sqrt{1-V^{2}}\cos \gamma -V\sin \gamma
\right) ^{2}\right] \log \left[  \left(\sqrt{1-V^{2}}\cos \gamma
-V\sin \gamma \right) ^{2}\right] \notag\\
&&+\frac{1}{2}\left[ \left( \sqrt{1-V^{2}}\sin \gamma +V\cos \gamma
\right) ^{2}\right] \log \left[ \left(
\sqrt{1-V^{2}}\sin \gamma +V\cos \gamma \right) ^{2}\right] \notag\\
&&-\frac{1}{2}\left[ \left( \sqrt{1-V^{2}}\sin \gamma +V\cos \gamma
\right) ^{2}+ \cos ^{2}\gamma \right] \log \left[  \left(
\sqrt{1-V^{2}}\sin \gamma +V\cos \gamma
\right) ^{2}+ \cos ^{2}\gamma \right] \notag\\
&&-\frac{1}{2}\left[ \left( \sqrt{1-V^{2}}\cos \gamma -V\sin \gamma
\right) ^{2}+ \sin ^{2}\gamma \right] \log \left[  \left(
\sqrt{1-V^{2}}\cos \gamma -V\sin \gamma \right) ^{2}+ \sin
^{2}\gamma \right].
\end{eqnarray}
\end{widetext}
We find that $S(\rho ^{Q})- S(\rho ^{Q}|\{{\Pi _{k}\}})$ is a
periodic function of the angle $\gamma$, and its cycle is $\pi/2$.
So, in order to obtain its maximum value, we just select the
appropriate value of $\gamma$ in the range of $0\leq\gamma\leq\pi/2$
for a given $V$. In principle, it is difficult to derive the
analytical expression of the maximum value because it involves a
transcendental equation. Here, we adopt a more rudimentary and
primitive method. As $S(\rho ^{Q})- S(\rho ^{Q}|\{{\Pi _{k}\}})$ is
a function of both the angle $\gamma$ and the visibility $V$, we can
fix one parameter $V$ and then find what value of $\gamma$ maximizes
$S(\rho ^{Q})- S(\rho ^{Q}|\{{\Pi _{k}\}})$. So, the corresponding
relationship between the parameters $\gamma$ and $V$ can be
revealed. It can be found that the angle $\gamma$ is a inverse
trigonometric function about $V$, i.e.,
\begin{eqnarray}
\label{III-10}
\gamma=\arcsin\left[\sqrt{\frac{1+\sqrt{1-V^{2}}}{2}}\right].
\end{eqnarray}
Substituting this value of $\gamma$ into Eq.~(\ref {III-09}), we can
obtain the CC
\begin{eqnarray}
\label{III-11} \mathcal{J}(\rho^{\prime
\prime})&=&\frac{1+\sqrt{1-V^{2}}}{2}\log
\left( 1+\sqrt{1-V^{2}}\right)  \notag \\
&&+\frac{1-\sqrt{1-V^{2}}}{2}\log \left( 1-\sqrt{1-V^{2}}\right),
\end{eqnarray}
which is shown by the red-dashed curve in Fig.~\ref{fig2.eps}.

From the Fig.~\ref{fig2.eps}, it can be observed that when $V=1$,
$CC=0$, and when $V=0$, $CC=1$. For the former case, since the two
pointer states of the WWD are the same, i.e., $\left\vert
d\right\rangle=U\left\vert d\right\rangle$, we cannot distinguish
the two paths of the photon by discriminating the two pointer
states. Moreover, both classical correlations and distinguishability
are equal to zero, i.e., $CC=D=0$, while the fringe visibility
$V=1$, which means that the wave-like behavior of the photon can be
perfectly observed. For the latter case, when two pointer states
$\left\vert d\right\rangle $ and $U\left\vert d\right\rangle $ are
mutually orthogonal, the two paths of the photon can be
distinguished, and both classical correlations and
distinguishability reach the maximum values, i.e., $CC=D=1$.
However, the wave-like behavior of the photon will disappear, i.e.,
$V=0$. When $0<V<1$, the photon is in a superposition. As the fringe
visibility $V$ increases, the CC decays monotonously, while the
distinguishability satisfies $D=\sqrt{1-V^{2}}$, which decays
monotonously with the visibility $V$ as well. Actually, since the
two pointer states of the WWD overlap, we can just partly
distinguish the two paths of the photon, while an imperfect
interference fringe is exhibited. According to the above analysis,
like the distinguishability, to some extent, the CC can be also used
to describe the which-way information of the photon. The more we
obtain the which-way information (the CC), the smaller the fringe
visibility $V$ will be. Furthermore, Eq.~(\ref{III-06},
\ref{III-11}) provide the complementary relation between the fringe
visibility and the CC. Indeed, the definitions of the
distinguishability and CC have a common basic--orthogonal
projection. We note that the definition of distinguishability is
related to the orthogonal projection vectors $\left\vert
M_{A}\right\rangle$ and $\left\vert M_{B}\right\rangle$ which are
expressed as in Eq.~(\ref{III-02}), while the definition of the CC
is related to the orthogonal projection vectors $\left\vert
M_{1}\right\rangle$ and $\left\vert M_{2}\right\rangle$ in
Eq.~(\ref{III-05}) with the condition
$\gamma=\arcsin\left[\sqrt{\frac{1+\sqrt{1-V^{2}}}{2}}\right]$.
After a simple calculation and contrast, we can find that the two
sets of orthogonal projections are the same. Then, it is natural to
use the CC to describe the which-way information of the photon. In
addition, because of the consistency between the CC and the
distinguishability, it is reasonable to conclude that the CC is just
the information gained about the particle-like property of the
photon, rather than the information gained about the wave-like
property of the photon.

\begin{figure}[tbp]
\includegraphics[clip=true,height=4.5cm,width=7.8cm]{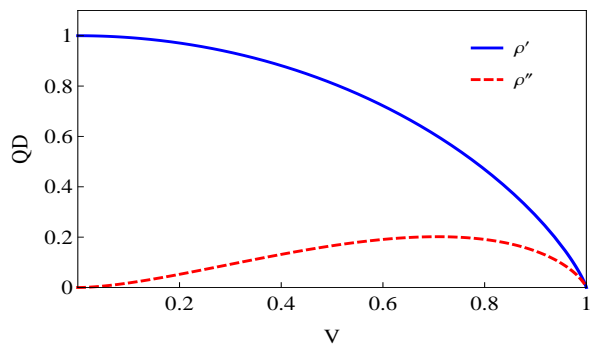}
\caption{(Color online). The relationship between the quantum
correlations and the fringe visibility $V$. } \label{fig3.eps}
\end{figure}
\subsection{Quantum correlations versus interference}

Quantum discord (QD) was originally introduced as an
information-theoretic approach to describe decoherence mechanisms in
a quantum measurement process. The wave-like property (the
visibility of the interference patten) of the photon is also related
to quantum coherence. Thus, there is should be a relation between
these, which will be considered below. Quantum discord,
$\mathcal{D}$, can be used to measure quantum correlations in a
bipartite system, and it is defined as the difference between the
total correlations, $\mathcal{I}$, and the classical correlations,
$\mathcal{J}$,
\begin{eqnarray}
\label{III-12} \mathcal{D}\left(\rho \right)= \mathcal{I}\left(\rho
\right)-\mathcal{J}\left(\rho \right).
\end{eqnarray}
Here, the total correlations is equal to quantum mutual information
\begin{eqnarray}
\label{III-13} \mathcal{I}(\rho)=S(\rho ^{Q})+S(\rho ^{D})-S(\rho ),
\end{eqnarray}
where $S$ is the von Neumann entropy, and $\rho^{Q}~(\rho^{D})$ is
the reduced density matrix of the photon (WWD).

(1) \emph{Based on von Neumann's notion of quantum measurement}

Considering the combined photon-detector system is in the entangled
state in Eq.~(\ref{III-01}), we can obtain the quantum correlations
between the photon and the WWD
\begin{eqnarray}
\label{III-13} \mathcal{D}\left( \rho^{\prime}\right)
&=&-\frac{1+V}{2}\log \left(\frac{1+V}{2}\right)-\frac{1-V}{2}\log
\left( \frac{1-V}{2}\right),
\end{eqnarray}
which is shown by the blue-solid curve in Fig.~\ref{fig3.eps}. We
can find that the quantum correlations are equal to the classical
correlations. In fact, for any pure state, the total correlations
are equally divided into the classical and quantum
correlations~\cite{Luo2008c}. In this case, the quantum correlations
has the same behaviors with the classical correlations.

(2) \emph{Based on Zurek's triple model of quantum measurement}

Considering the combined photon-detector system is in the state
$\rho^{\prime\prime}$ in Eq.~(\ref{III-07}), the QD between the
photon and the WWD can be written as
\begin{eqnarray}
\label{III-14} \mathcal{D}\left(\rho^{\prime\prime}\right)
&=&-\frac{1+V}{2}
\log \left(\frac{1+V}{2}\right) -\frac{1-V}{2}\log \left( \frac{1-V}{2}\right)  \notag\\
&&-\frac{1+\sqrt{1-V^{2}}}{2}\log \left( 1+\sqrt{1-V^{2}}\right)  \notag\\
&&-\frac{1-\sqrt{1-V^{2}}}{2}\log \left( 1-\sqrt{1-V^{2}}\right).
\end{eqnarray}
We plot the QD as a function of the fringe visibility $V$ as shown
in the red-dashed curve in Fig.~\ref{fig3.eps}. It can be seen that
the QD is not a monotonic function of the fringe visibility $V$.
When $V=0$, we have QD $=0$, and CC $=1$. Since the two pointer
states of the WWD are mutually orthogonal, a perfect discrimination
among the two paths of the photon can be achieved. From the point of
view of quantum information theory, there are only classical
correlations and no quantum correlations between photon and WWD, and
such state of the combined photon-detector system is a classical
state~\cite{Luo2008a,Luo2008b}. Here, we note that when one make a
perfect quantum measurement, the state of the combined
photon-detector system should be a classical state. In this case, we
can confirm that no quantum correlations correspond to no fringe
visibility. As $V$ increases, the QD first increases and then
decreases. In this process, since the two pointer states of the WWD
have an overlap, the state of the combined photon-detector system
becomes a separable state~\cite{Werner1989}. Although there is no
entanglement in a separable state, some classical correlations (CC)
and quantum correlations (QD) may exist. Here, we can see that CC is
used to describe the which-way information, but QD is different from
the fringe visibility. When $V=1$, both CC and QD are zero. Since
the two pointer states of the WWD are the same, from
Eq.~(\ref{III-07}) the state of the combined photon-detector system
is changed to a product state or an uncorrelated state. In this
case, the QD is equal to zero; however, the fringe visibility $V=1$.
Obviously, the QD is different from the fringe visibility $V$,
though both of these are related to quantum coherence.


\section{\label{Sec:4} discussion and conclusion}


In the quantitative relation formulation of wave-particle duality,
the wave nature is described by the visibility of the interference
pattern, while the particle nature is characterized by the path
distinguishability, which is based on von Neumann's measurement
theory. In modern measurement theory, to acquire the which-way
information about a particle traveling, a ``triple model" (which
consists of a measured system, an apparatus and environment) must be
introduced. Here, we use Zurek's ``triple model" to study
wave-particle duality in a symmetric Mach-Zehnder interferometer,
where the interaction between the quantum system and the apparatus
produces a quantum entanglement between them, and later the coupling
of the environment and the apparatus generates a triple entanglement
among the system, apparatus and environment. By tracing the
environment, the state of system and apparatus is no longer an
entangled state, but a correlated state. With the help of quantum
information theory, it is easy to see that the correlations between
the measured system and detector [which include classical
correlations (CC) and quantum correlations (QD)] are related to the
information gain about the measured system. It is found that the CC
is the information gain about the particle-like property of the
measured system which is consistent with the path
distinguishability. Moreover, we can also see that QD and the
visibility have different values, but both of them represent one
single phenomenon--quantum coherence. Finally, we derive an
analytical expression for the QD of one type of two-qubit separable
states, i.e., quantum-classical states in Eq.~(\ref {III-08}). Since
the analytical expression for the QD can be written only for some
special type of two-qubit states at present~\cite{Luo2008c}, our
analysis broadens the regime of analytical expressions for the QD of
two-qubit states.

\begin{acknowledgments}
This work is supported by the National Natural Science Foundation of
China under Grants No. 11374095, No. 11422540, No. 11434011, No.
11575058 and the National Fundamental Research Program of China (the
973 Program) under Grants No. 2012CB922103.
\end{acknowledgments}

\end{document}